\documentclass[fleqn,usenatbib]{mnras}

\usepackage[T1]{fontenc}
\usepackage{ae,aecompl}

\usepackage{hyperref}
\usepackage{booktabs}
\usepackage{multirow}
\usepackage{pgf}
\usepackage{subcaption}
\usepackage{threeparttable}

\usepackage{graphicx}	
\usepackage{amsmath}	
\usepackage{amssymb}	

\usepackage{newtxtext,newtxmath}

\newcommand{\DPS}{${\Delta\mbox{PS}}$}

\newcommand{\Ha}{\ion{H}{$\;\!\!\alpha$}}
\newcommand{\Hb}{\ion{H}{$\;\!\!\beta$}}
\newcommand{\Oiii}{[\ion{O}{III}]}

\newcommand{\Nii}{[\ion{N}{II}]}
\newcommand{\Sii}{[\ion{S}{ii}]}

\newcommand{\BPT}{\mbox{BPT}}

\defcitealias{Tous+2020}{TSP20}

\title[Activity diagnostic via PCA of visible spectra]{Fully comprehensive diagnostic of galaxy activity using principal components of visible spectra: implementation on nearby S0s}

\author[J.\ L.\ Tous et al.]{
J.\ L.\ Tous$^{1,2,3}$\thanks{E-mail: j.l.tous-mayol@soton.ac.uk}, J.\ M.\ Solanes$^{2,3}$ and J.\ D.\ Perea$^4$\\
$^{1}$School of Physics and Astronomy, University of Southampton, Highfield, Southampton SO17 1BJ, UK\\
$^{2}$Institut de Ci\`encies del Cosmos (ICCUB), Universitat de Barcelona, Mart\'i i Franqu\`es 1, E-08028 Barcelona, Spain\\
$^{3}$Departament de F\'isica Qu\`antica i Astrof\'isica, Universitat de Barcelona, Mart\'i i Franqu\`es 1, E-08028 Barcelona, Spain\\
$^{4}$Departamento de Astronom\'ia Extragal\'actica, Instituto de Astrof\'isica de Andaluc\'ia, IAA-CSIC, Glorieta de la Astronom\'ia s/n, E-18008 Granada, Spain\\
}
\date{Accepted XXX. Received YYY; in original form ZZZ}

\pubyear{2024}

\begin{document}
\label{firstpage}
\pagerange{\pageref{firstpage}--\pageref{lastpage}}
\maketitle

\begin{abstract}
We introduce a novel galaxy classification methodology based on the visible spectra of a sample of over $68,000$ nearby ($z\leq 0.1$) Sloan Digital Sky Survey lenticular (S0) galaxies. Unlike traditional diagnostic diagrams, which rely on a limited set of emission lines and class dividers to identify ionizing sources, our approach provides a comprehensive framework for characterizing galaxies regardless of their activity level. By projecting galaxies into the 2D latent space defined by the first three principal components (PCs) of their entire visible spectra, our method remains robust even when data from individual emission lines are missing. We employ Gaussian kernel density estimates of the classical Baldwin-Phillips-Terlevich (BPT) activity classes in the new classification subspace, adjusted according to their relative abundance in our S0 sample, to generate probability maps for star-forming, Seyfert, composite, and LINER galaxies. These maps closely mirror the canonical distribution of BPT classes shown by the entire galaxy population, demonstrating that our PC-based taxonomy effectively predicts the dominant ionizing mechanisms through a probabilistic approach that provides a realistic reflection of galaxy activity and allows for refined class membership. Our analysis further reveals that flux-limited BPT-like diagrams are inherently biased against composite and star-forming galaxies due to their weaker \Oiii\ emission. Besides, it suggests that although most low-activity galaxies excluded from these diagnostics exhibit visual spectra with LINER-like characteristics, their remaining activity is likely driven by mechanisms unrelated to either star formation or supermassive black hole accretion. A machine-readable catalogue listing BPT-class probabilities for the galaxies analysed is available online at CDS website.  
\end{abstract}


\section{Introduction}
\label{S:intro}

The growth of central supermassive black holes (SMBH) and the regulation of star formation (SF) in galaxies are two closely intertwined astrophysical processes. Energy released by active galactic nuclei (AGN), when absorbed by the host galaxy, can suppress SF in massive galaxies by heating and expelling ambient gas, a phenomenon known as (negative) AGN feedback. Conversely, AGN outflows can also trigger in some cases SF by compressing dense gas clouds, a positive feedback mechanism more likely to occur in low-luminosity objects. The interplay between nuclear black hole growth and SF is pivotal in explaining the evolution of galaxies over cosmic time \citep[see, e.g.][and references therein]{Mulcahey+2022}. Without AGN feedback, cosmological models of galaxy formation and evolution fail to accurately reproduce both the observed properties of galaxies and their stellar mass and luminosity functions. However, gaining a comprehensive understanding of AGN accretion and SF requires reliably identifying these processes when they occur, whether independently or simultaneously, within galaxies. 

Most traditional diagnostic diagrams of activity in galaxies -- whether linked to nuclear black hole accretion or normal stellar processes -- rely on spectroscopy to identify ionizing sources. Although nearly all windows of the electromagnetic spectrum have been employed with this purpose (e.g. \citealt{Condon1992}; \citealt{Kennicutt+1998}; \citealt{Ranalli+2003}; \citealt{Best+05}; \citealt{Cid+2010}; \citealt{Stern+2012}, \citealt{Agostino+2019}) --, the Baldwin-Phillips-Terlevich (BPT; \citealt{Baldwin+1981}) classification scheme remains one of the most widely used. Among its various forms, the BPT-\ion{N}{ii} diagnostic diagram\footnote{Hereafter referred to simply as the BPT diagram.} stands out as the most popular. This version uses the \Oiii$\lambda5007$/\Hb\ and \Nii$\lambda 6584 /$\Ha\ flux ratios to effectively distinguish between galaxies whose line emission primarily originates from \ion{H}{ii} regions, i.e.\ it is basically associated with SF, and those that also or exclusively exhibit either high-ionization regions (classified as AGNs or Seyferts) or low-ionization nuclear emission-line regions (LINERs). By plotting a sample of approximately one hundred thousand galaxy spectra from the Sloan Digital Sky Survey \citep{York+2000} onto the BPT diagram, \citet{Kauffmann+2003} demonstrated that star-forming galaxies and AGN-dominated systems occupy in this subspace two distinct, wing-shaped areas that converge at the base, giving the distribution of data points its characteristic 'seagull' shape. To more precisely define the zones of the BPT diagram corresponding to different activity classes, various model-based and empirical dividing lines have been proposed \citep[e.g.][]{Kewley+2001, Kauffmann+2003, Stasinska+2006, Kewley+2006, Cid+2010}. These dividers facilitate the selection of the different types of active galaxies, as long as the four emission lines involved in the classification are present in their spectra with an acceptable signal-to-noise ratio (S/N). 

Another widely used classification scheme for emission-line galaxies is the WHAN diagram introduced by \citet{Cid+2010}, which streamlines the classification of galactic activity by requiring only measurements from two spectral lines. This diagram utilizes the \Ha\ equivalent width (EW) and the \Nii$\lambda 6584 /$\Ha\ flux ratio to differentiate between star-forming galaxies, strong AGNs, weak AGNs or true LINERs, and a newly identified class of 'retired' galaxies which, according to these authors, are objects whose weak nebular emission is driven by the ionizing radiation of old post-AGB stars. More recently, \citet{Sanchez+2024} have proposed the WHaD diagram, an even simpler approach that relies on the EW(\Ha) and the velocity dispersion of this same line to classify ionizing sources, also defining new regions to differentiate between them.

Apart from their dependence on measurable emission lines, another important limitation of all these classification schemes is their reliance on discrete class definitions. While the class boundaries are based on sound physical principles, they do not trace real discontinuities in the distributions of measurements within the diagrams. Although the different physical processes driving galaxy activity map to distinct regions, galaxies located in adjacent locations are expected to have similar spectra and ionization sources, even when positioned on opposite sides of a class divider. Additionally, it is not uncommon for SF and nuclear black hole accretion to coexist within the same galaxy (e.g.\ \citealt{Filho+2004}), with both contributing comparably to its line emission, making the assignment of a single activity class unsuitable for such systems. In fact, the introduction of a composite class in the BPT diagram -- encompassing approximately the central region in between the two wings -- stems from the inherent ambiguity in the classification of objects whose activity is driven by both physical mechanisms \citetext{see, e.g.\ the detailed discussion in Sec.\ 5.1 of \citealt{Cid+2010} regarding the difficulties in correctly identifying star-forming and AGN galaxies in this diagram}. A second major limitation of galaxy classification schemes based on emission lines is their inherent exclusion of fully quiescent objects with lineless spectra. Moreover, diagnostic diagrams that rely on a limited set of narrow emission lines, even those using just one or two lines, often struggle to provide reliable classifications for galaxies with spectral features that do not reach a minimum S/N value.

This study continues a series of works investigating the activity of present-day lenticular (S0) galaxies through the analysis of variance in their full visible spectra. As shown in \citeauthor{Tous+2020} (\citeyear{Tous+2020}; hereafter \citetalias{Tous+2020}), the first two principal components, PC1 and PC2, derived from these spectra provide an effective means of classifying activity within this Hubble type. The projections of the spectra onto the plane defined by these components\footnote{For convenience, the coefficients obtained by projecting individual spectra onto the orthogonal eigenspectra will be also referred to as PC1, PC2, and so on.} reveal a 2D distribution characterized by two prominent features (see Fig.~\ref{fig:pc1_pc2} in the Appendix). On the left-hand side there is a narrow, negatively inclined, lens-shaped region, termed the passive sequence (PS), which encapsulates approximately $70$ per cent of nearby S0s and serves as the locus of quiescent galaxies with lineless spectra. Extending to the right of this compact region is a more dispersed domain known as the active cloud (AC), which includes spectra from star-forming (mostly) and AGN systems, together representing nearly $25$ per cent of nearby S0 galaxies. The remaining small fraction of S0s are objects with intermediate spectral properties that define a thin transition region (TR) separating these two main areas.

As shown in \citet{JimP+22}, the distance from the PS ridge given by the equation
\begin{equation}
    \Delta\mbox{PS} = \log\left[2.913+0.758\cdot\mbox{PC1}+0.652\cdot\mbox{PC2}\right]\;,
    \label{eq:dps}
\end{equation}
serves as a robust indicator of the activity level in S0 galaxies. This metric is so effective that the spectral classes PS, TR, and AC, defined by different ranges of values of equation~\eqref{eq:dps}, align precisely with the EW(\Ha) dividers of the WHAN diagram proposed by \citet{Cid+2010}. Further applications of this metric to data cubes from the Mapping Nearby Galaxies at Apache Point Observatory \citep[MaNGA;][]{Bundy+2015} survey, aimed at studying the radial distribution of activity in the disks of S0 galaxies, are presented in \citet{Tous+2023, Tous+2024}.

Now, we seek to delve deeper into the spectral classification of S0 galaxies by incorporating the third principal component (PC3) of their visible spectra to enable a more precise identification of the primary ionization source within these systems. The manuscript is organized as follows: 

Section~\ref{S:sample} details the selection of the galaxy sample and the input data used. In Section~\ref{S:activity_variance}, we outline our methodology for determining galaxy activity based on the first three principal components (PCs) of their visible spectra. Section~\ref{S:classification} introduces a new classification scheme that allows membership in a given activity class to be determined through probabilistic criteria instead of rigid class boundaries for galaxies with any level of activity, provided their spectra meet minimal quality standards. The implications and insights derived from this methodology are discussed in Section~\ref{S:discussion}, with the conclusions summarized in Section~\ref{S:conclusion}. While this study focuses on present-day S0 galaxies, future work will extend this classification paradigm to the entire Hubble sequence.

\section{Sample selection and data processing}
\label{S:sample}
In this work, we use the same sample of $68,043$ S0 galaxies analysed in \citetalias{Tous+2020}. These are objects with apparent Petrosian $r$-band magnitudes $\leq 17.77$ and heliocentric redshifts $0.01 \leq z \leq 0.10$ retrieved from the $12$th data release of the Sloan Digital Sky Survey \citep[SDSS;][]{Alam+15}. Their lenticular morphology is confirmed by meeting the criteria $T \leq 0$ and $P_{\rm S0} > 0.7$, where $T$ is the numerical Hubble stage and $P_{\rm S0}$ is the probability of being an S0 galaxy, as listed in the morphological catalogue of SDSS galaxies by \citet{HDS+2018}.

\begin{figure*}
    \centering
        \includegraphics[width=\textwidth]{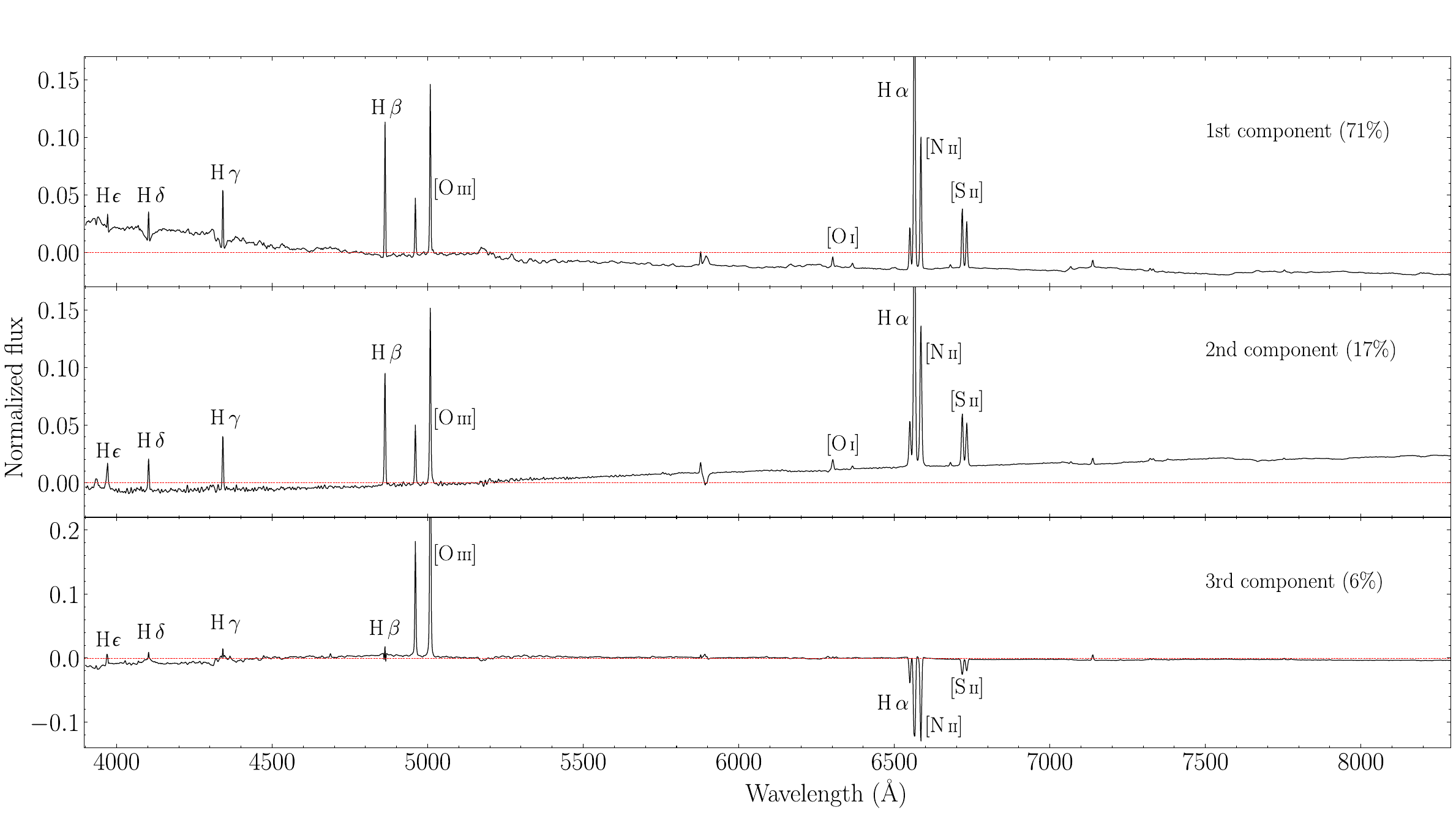}
    \caption{First three eigenspectra of the sample of present-day S0 galaxies from the SDSS analyzed in \citetalias{Tous+2020}. The percentages in parentheses represent the proportion of total variance explained by each component. A horizontal red dashed line indicates the zero level for each PC. The most important spectral features of the PCs are labelled. Note that the actual emission or absorption nature of these lines in real spectra is determined by the sign of the weight factor associated with each PC.}
    \label{fig:PCs}
\end{figure*}

The spectroscopic data consist of single-fibre visible spectra obtained by the SDSS spectrograph \citep{Smee+13}, comprising nearly $4000$ flux bins with wavelength coverage from $3800$ to $9200$ \AA\ and an average spectral resolution of $R \sim 2000$. The original spectra are corrected for Galactic dust reddening using a standard \citet{Fitzpatrick+1999} dust extinction model and shifted to the rest-frame, maintaining a pixel spacing for the flux vectors of $10^{-4}$ dex. The spectra are then normalized using equation (3) from \citetalias{Tous+2020} and their dimensionality reduced to the projections onto their first PCs\footnote{The first five principal components, including the three used in this work, are publicly available in \citet{JimP+22_table}}. The present analysis focuses on the first three PCs, with the first two combined into a single measurement through equation \eqref{eq:dps}. This spectral information is complemented with emission line properties retrieved from the Portsmouth Stellar Kinematics and Emission Line Fluxes by \citet{Thomas+2013}. The reader is directed to \citetalias{Tous+2020} for a detailed explanation of the processing of the spectral data.

\section{Activity as a source of spectral variance}
\label{S:activity_variance}
The intrinsic variability in the spectra of most galaxies is primarily driven by SF and/or nuclear emission processes. As demonstrated in our previous works, much of this variability in present-day S0s is captured by the first two PCs, which, as shown in Fig.~\ref{fig:PCs}, encompass key spectral features such as the Balmer series and prominent forbidden emission lines like \Sii, \Nii, and \Oiii.

\begin{figure}
    \centering
	\includegraphics[width=\columnwidth]{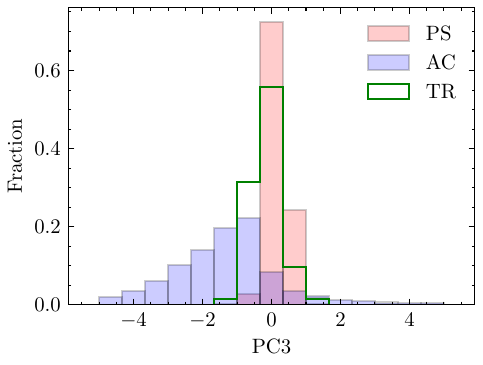}
    \caption{Distribution of PC3 values for the different spectral classes of nearby S0 galaxies: passive (PS, red), active (AC, blue), and transition (TR, green). This component accounts for six percent of the total sample variance.}
    \label{fig:pc3}
\end{figure}

However, these two components alone cannot properly distinguish the distinct spectral signatures produced by the various ionizing mechanisms at work in galaxies. To capture this additional variance, it is essential to include more spectral information. This need is emphasized in Fig.~\ref{fig:pc3}, which displays the distributions of the values of (the coefficients associated with) PC3 across the three spectral classes of S0 galaxies. The relatively narrow distributions shown by the quiescent PS and mildly active TR S0s suggest that the primary physical processes influencing their spectra are largely captured by the first two components. In contrast, the third component exhibits significantly greater variability in lenticular galaxies of the AC class, highlighting the importance of incorporating this factor to enhance the classification's discriminating power. Thus, it becomes evident that the differentiation between sources where SF provides the ionizing photons and those requiring a harder ionizing spectrum should become much more apparent in a '\DPS-PC3' diagram, where \DPS\ -- the parameter defining the three spectral classes of S0 galaxies -- is plotted against PC3.

\begin{figure}
    \centering
	\includegraphics[width=\columnwidth]{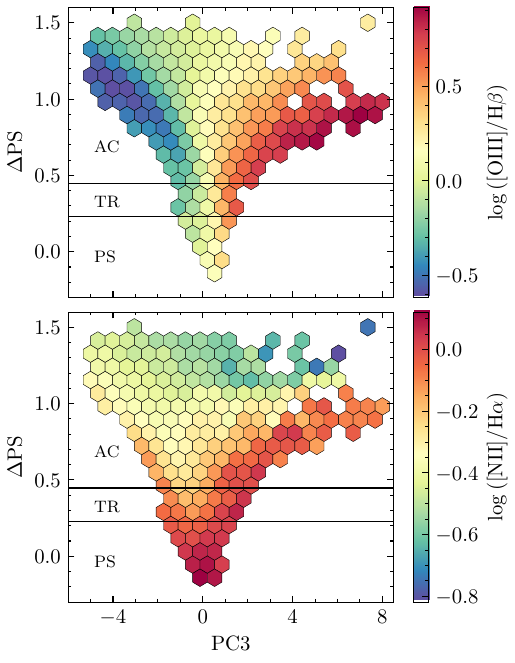}
    \caption{Distribution of \Oiii/\Hb\ (top) and \Nii/\Ha\ (bottom) flux ratios in the \DPS-PC3 plane. The colours indicate the median value of these flux ratios within each hexagonal cell, calculated for S0 galaxies whose spectra have an amplitude-over-noise (AoN) ratio greater than 2 in the corresponding emission lines. The horizontal dividers, at \DPS\ values of $0.23$ and $0.44$ mark the boundaries between the AC, TR and PS regions, as defined by \citet{JimP+22}. Complementary colour maps showing the distributions of the EW(\Ha) and EW(\Oiii) can be seen in Fig.~\ref{fig:dps_pc3_ssfr_oiii} in the Appendix.}
    \label{fig:dps_pc3_bpx_bpy}
\end{figure}

As shown in the bottom panel of Fig.~\ref{fig:PCs}, the eigenspectrum PC3 is primarily characterized by the presence of a strong \Oiii\ doublet in emission, with weaker absorption \Ha, \Nii, and \Sii\ lines, on top of a flat continuum. Therefore, for any spectrum, this component regulates the strength of the oxygen doublet with respect to the other emission lines once the lower order PCs are fixed. Compared to \ion{H}{II}\ regions, AGN have higher \Oiii$\lambda5007$/\Hb\ and \Nii$\lambda 6584 /$\Ha\ flux ratios because accretion processes produce high-energy photons that heat the narrow-line region, resulting in stronger collisionally excited metallic lines relative to the Balmer recombination lines \citep{Stasinska+2006}. In the colour maps of Fig.~\ref{fig:dps_pc3_bpx_bpy}, we show the median values of these flux ratios across the 'V'-shaped region outlined by the spectral data in the \DPS-PC3 plane. In the top panel, the median \Oiii$\lambda5007$/\Hb\ flux ratio in the AC sector of the diagram increases monotonically along PC3, with the left and right edges of the 'V' corresponding to the spectra with the lowest and highest ratio values, respectively, while the PS and TR sectors exhibit intermediate values. Conversely, the bottom panel of Fig.~\ref{fig:dps_pc3_bpx_bpy} shows that while the distribution of \Nii$\lambda 6584$/\Ha\ flux ratios also peaks at the right edge of the AC sector, these maxima extend downwards into the TR and PS sectors, converging in the lower vertex of the diagram near the origin of coordinates. In contrast, the left edge of the diagram now displays intermediate flux ratios, with the lowest values found in the upper right region. The inverse mapping, showing the distributions of PC3 and \DPS\ across the 'two-wing'-shaped BPT diagram, can be found in Fig.~\ref{fig:bpt_cmap} in the Appendix. These findings underscore the effectiveness of combining the parameters \DPS\ and PC3 to disentangle the different ionization sources in lenticular galaxies. 

\begin{figure*}
    \centering
	\includegraphics[width=\textwidth]{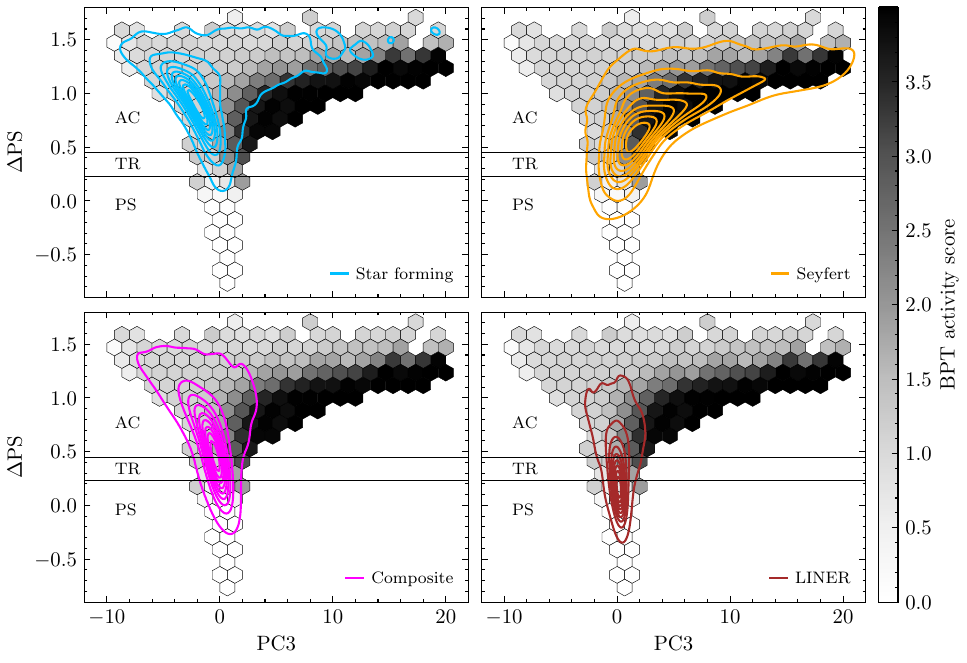}
    \caption{Kernel density estimates of the distribution of S0 galaxies across BPT classes star-forming, Seyfert, composite, and LINER in the \DPS-PC3 diagram. The hexagonal bins in the background, consistent across all panels, show the distribution of the entire S0 sample in this latent space, with greyscale intensity indicating the BPT activity score of the galaxies in each cell, as defined by equation~\eqref{eq:a_score}, which reflects the average BPT class at the corresponding location (see Sec.~\ref{SS:p_bpt}). The horizontal lines in all panels are the same \DPS\ demarcations defined in Fig.~\ref{fig:dps_pc3_bpx_bpy}. The distribution of galaxies lacking a BPT classification is depicted in Fig.~\ref{fig:dps_pc3_blank}.}
    \label{fig:dps_pc3_bpt}
\end{figure*}

To further validate the diagnostic power of the \DPS-PC3 diagram, Fig.~\ref{fig:dps_pc3_bpt} presents the kernel density estimates in this plane of the distributions for $9907$ Star forming, $1376$ Seyfert, $8327$ Composite, and $3196$ LINER galaxies in our S0 sample, displayed from top-left to bottom-right, respectively. These distributions are based on BPT classifications following the demarcation lines defined by \citet{Kauffmann+2003}, \citet{Kewley+2001}, and \citet{Cid+2010}. For this analysis, only spectra with an amplitude-over-noise ratio (AoN) larger than 2 in the four emission lines used for classification (i.e.\ \Ha, \Hb, \Nii, and \Oiii) have been considered.\footnote{AoN is a proxy for S/N;  therefore, the terms will be used interchangeably in this paper.} The results show that star-forming lenticular galaxies are predominantly aligned along the left edge of the AC sector in the diagram, with some spread into its central and upper regions. Seyferts, on the other hand, dominate the right-hand side of the diagram, primarily occupying the AC sector but also extending into the lower TR and PS sectors. For their part, the distributions of composite and LINER S0s are centred, respectively, just above and below the boundaries of the TR sector, with LINERs exhibiting a particularly narrow range of PC3 values centred around zero. The plots also reveal that the lower tip of the \DPS-PC3 diagram, which extends deep into the PS sector, consists of galaxies whose spectra are considered unclassifiable in a BPT representation. These galaxies are discussed in detail in Sec.~\ref{SS:unclass}.

\section{Galaxy classification using the principal components of the visible spectrum}
\label{S:classification}
In the previous section, we have demonstrated that the first three PCs of the visible spectrum provide a robust classification for present-day S0 galaxies. Building on this, we now emphasize the strong discriminating power of the \DPS-PC3 diagram and its ability to classify the entire S0 population, including objects that are typically excluded from traditional emission-line diagrams due to the absence or unreliability of some or all diagnostic line measurements.

\subsection{Probability that a galaxy belongs to a BPT class}
\label{SS:p_bpt}
The probability that a galaxy in our present-day S0 sample is of class $\BPT_{i=0,...,4}=$ [unclassified, star-forming, composite, LINER, Seyfert], given its \DPS\ and PC3   coordinates, can be computed as follows,
\begin{equation}
    P(\BPT_i|{\rm PC3, \Delta PS}) = \dfrac{P({\rm PC3, \Delta PS}|\BPT_i) P(\BPT_i)}{\sum_{i=0}^4 P({\rm PC3, \Delta PS}|\BPT_i) P(\BPT_i)}\;.
    \label{eq:p_bpt}
\end{equation}
In this equation, $P({\rm PC3, \Delta PS}|\BPT_i)$ is derived from the Gaussian kernel density estimate for each BPT class, using a bandwidth of $n_i^{-1/6}$, where $n_i$ denotes the number of galaxies in the $i$th class \citep{Scott2015}, and $P(\BPT_i)$ represents the fraction of galaxies in the sample belonging to class $\BPT_i$. In our dataset these fractions are $P(\BPT_0) = 0.66$, $P(\BPT_1) = 0.15$, $P(\BPT_2) = 0.12$, $P(\BPT_3) = 0.05$, and $P(\BPT_4) = 0.02$.

From these probabilities one can derive a BPT activity score that determines the average contribution of the BPT classes across the different regions of the \DPS-PC3 diagram:
\begin{equation}
\overline{BPT}=\sum_{i=0}^4 i\cdot P(\BPT_i | {\rm PC3, \Delta PS})\;.
    \label{eq:a_score}
\end{equation}
As shown in Fig~\ref{fig:dps_pc3_bpt}, the BPT activity score values, averaged within hexagonal cells shaded in varying tones of gray, align closely with the density distributions of the various BPT classes, represented by colour-coded density contour lines drawn on top of the \DPS-PC3 diagram. Note that the BPT-scores we have just defined can help identify the BPT class most likely contributing to a galaxy's classification, but they should not be used to determine its primary BPT type. For that purpose, one should refer to the $BPT_i$ probabilities derived from eq.~\eqref{eq:p_bpt} assuming $P({\rm BPT}_0) = 0$. These probabilities are provided in Table~\ref{tab:catalogue} included in the Appendix for all galaxies analysed in this work.

\begin{figure}
    \centering
	\includegraphics[width=\columnwidth]{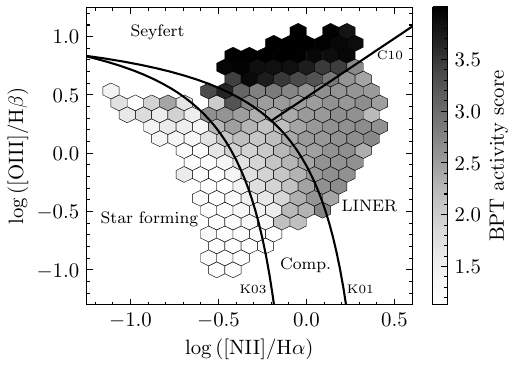}
    \caption{Distribution of BPT activity scores for present-day S0 galaxies from the SDSS, averaged within hexagonal bins. Note that in this diagram, the lowest possible score is $\overline{BPT}=1$, corresponding to pure star-forming galaxies (see the text for details). Black curves represent standard class boundaries: the line labelled K01 is the demarcation proposed by \citet{Kewley+2001}, K03 by \citet{Kauffmann+2003}, and C10 by \citet{Cid+2010}. This figure includes only galaxies with spectra that have an AoN$\,> 2$ for all four diagnostic emission lines. The number of galaxies per cell corresponding to this BPT diagram is shown in Fig.~\ref{fig:bpt_hist} in the Appendix.}
    \label{fig:bpt_activity_score}
\end{figure}

Fig.~\ref{fig:bpt_activity_score} shows the distribution of BPT activity scores mapped onto the BPT diagram of our sample. These scores were computed using eqs.~\eqref{eq:p_bpt} and \eqref{eq:a_score}, under the condition $P(\BPT_0) = 0$, to exclude unclassified galaxies not represented in the diagram. Under this condition, $P(\BPT_i|{\rm PC3, \Delta PS})$ assigns each galaxy a probability of belonging to one of the four BPT classes based on the first three PCs of its visible spectrum. The close correspondence between the BPT activity score mapping and the different BPT regions (see Section~\ref{S:discussion} for explanations of the individual probability maps for each BPT class) highlights the effectiveness of these spectral components as predictors of galaxy activity. Furthermore, this probabilistic approach more accurately captures the complexity of nature compared to the traditional assignment of a single activity class, and it can be leveraged to enhance the purity of subsamples targeting specific BPT types.

\subsection{A classification scheme that goes beyond emission lines}
\label{SS:unclass}
As previously noted, the primary strength of our approach lies in its comprehensiveness. By utilizing the first three PCs derived from the entire visible spectrum, our method captures not only the active galaxies typically represented in BPT-like diagrams, but also those excluded due to missing or unreliable emission line data, whether caused by observational challenges (e.g.\ noisy spectra), technical limitations (e.g.\ issues with line fitting), or the galaxies' inherently nearly or fully quiescent nature. This is particularly important for galaxy datasets like the sample of S0s studied here, where $45,220$ objects have spectra with either an AoN$\,<2$ for relevant emission lines or missing or invalid entries in the Portsmouth catalogue, rendering them unclassifiable with conventional emission-line diagrams. 

\begin{figure}
    \centering
	\includegraphics[width=\columnwidth]{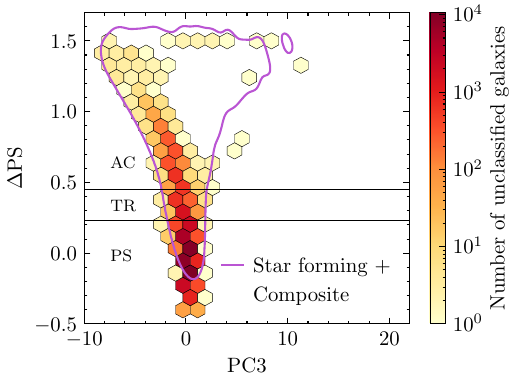}
    \caption{Distribution of S0 galaxies without a BPT classification due to spectra with an AoN$\,<2$ or invalid flux measurements in any of the BPT emission lines binned in hexagonal cells. Darker tones of colour represent higher data point densities. For reference, the purple contour outlines the $90$ per cent level of the kernel density estimate for the combined distribution of star-forming and composite S0 galaxies. The horizontal lines are the same \DPS\ demarcations defined in Fig.~\ref{fig:dps_pc3_bpx_bpy}.}
    \label{fig:dps_pc3_blank}
\end{figure}

Fig.~\ref{fig:dps_pc3_blank} displays the distribution of such unclassified galaxies in the \DPS-PC3 latent space. As noted by \citet{JimP+22}, most of them -- specifically, $40,580$ objects in the current sample -- belong to the PS spectral class. The members of this class are expected to eventually develop a very low-level, flat radial activity profile once all gas ionizing sources are extinguished \citep{Tous+2024}, justifying their exclusion from traditional emission-line diagnostics. The calculation for these objects of $P(\BPT_i|{\rm PC3, \Delta PS})$ using equation~\eqref{eq:p_bpt}, with $P({\rm BPT}_0) = 0$, reveals that $1726$ galaxies have a probability exceeding $0.5$ of being classified as composite, while $38,038$ exhibit the same likelihood for the LINER category, consistent with the distribution of BPT types observed in the \DPS--PC3 latent space (see Fig.~\ref{fig:dps_pc3_bpt}). The significant prevalence among the S0--PS of systems whose optical spectra have PCs that align predominantly with the LINER class suggests that any residual activity they may harbour\footnote{Less than $6.9\%\,(2999/43,578)$ of S0--PS galaxies have all four diagnostic lines with sufficient S/N strength to be included in the BPT diagram.} might not necessarily be driven by feeble nuclear accretion, but instead by hot post-AGB stars \citep{Cid+2010,Byler+2017}.

\begin{figure}
    \centering
	\includegraphics[width=\columnwidth]{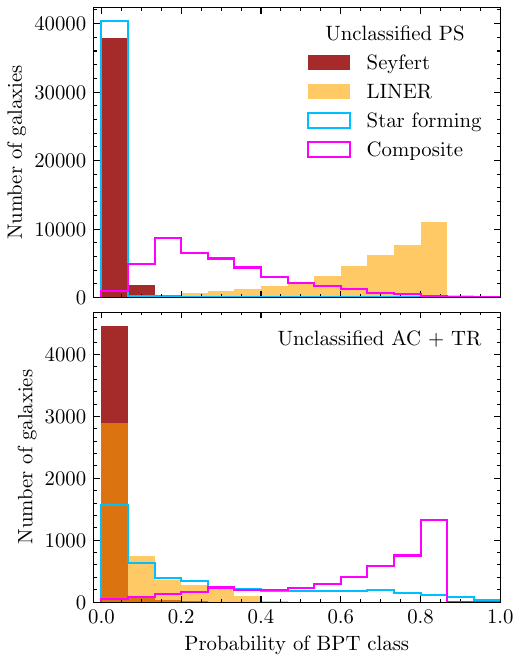}
    \caption{Number of S0 galaxies located in the PS (top), and TR+AC sectors (bottom) of the \DPS-PC3 diagram that lack a reliable traditional BPT classification, plotted against the probability of belonging to each BPT classes. Probabilities are calculated from the first three PCs of their entire visible spectra using equation~\eqref{eq:p_bpt}, with $P({\rm BPT}_0) = 0$}.
    \label{fig:p_bpt_unclass}
\end{figure}

A significant number of unclassified S0s are also scattered across a broad area of the diagram, extending from the TR sector to the left side of the AC region -- $2034$ and $2606$ objects, respectively --, where composite and star-forming BPT classes predominate. Indeed, as it is apparent from the histograms in the bottom panel of Fig~\ref{fig:p_bpt_unclass}, we find that up to $3489$ of these galaxies have a probability $> 0.5$ of being classified as composite in the BPT taxonomy, while $1011$ are similarly identified with star-forming systems. Conversely, most of these unclassified S0s exhibit low probabilities of belonging to the Seyfert or LINER classes, with typical values below $\sim 0.1$. For further discussion about the nature of all objects lacking a traditional BPT classification, see next section.

\section{Implications and insights from the new classification system}
\label{S:discussion}
\begin{figure*}
    \centering
	\includegraphics[width=\textwidth]{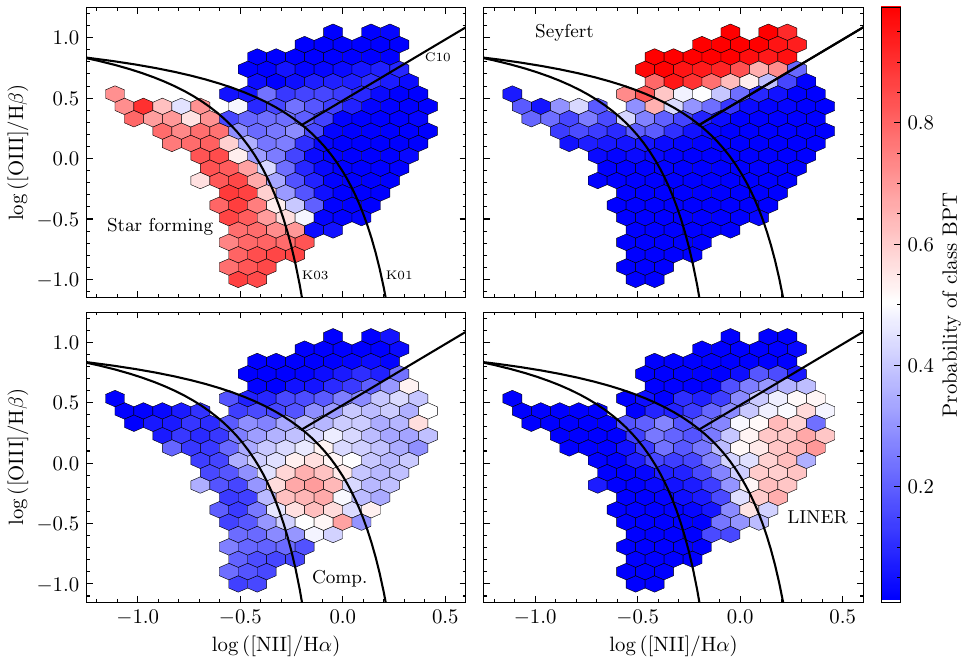}
    \caption{From top left to bottom right, the panels display on a colour scale the probabilities for present-day S0 galaxies from the SDSS to be classified as star-forming, Seyfert, composite, and LINER, as calculated using eq.~\eqref{eq:p_bpt}. As in  Fig.~\ref{fig:bpt_activity_score}, the probabilities were computed under the condition $P(\BPT_0) = 0$ to exclude galaxies lacking a BPT classification and their values averaged within hexagonal bins. The class dividers are the same as in Fig.~\ref{fig:bpt_activity_score}. Only galaxies with spectra having AoN$\,> 2$ in the diagnostic emission lines are included.}
    \label{fig:bpt_p_bpt}
\end{figure*}

As shown in Figs.~\ref{fig:bpt_activity_score} and \ref{fig:bpt_p_bpt}, the probability distributions for the star-forming, composite, LINER, and Seyfert classes of present-day S0s, inferred by mapping their \DPS-PC3 coordinates onto the BPT emission-line diagram, confirm that our PC-based classification scheme is a highly effective tool for predicting the dominant ionizing source in galaxies. The median probability of belonging to the corresponding BPT class exceeds $0.5$ in a good part of the four BPT regions, with the highest probabilities observed in broad areas of the Seyfert and star-forming sectors, reflecting the efficient segregation of these two types of sources in the \DPS-PC3 plane. In contrast, the probability distributions of composite and LINER S0s generally show lower values, as these two classes are more intermixed in the \DPS-PC3 latent space. Even so, in both cases, the zones where the median probabilities peak align precisely with the heart of the respective sectors.

The stochastic nature of this new approach to galaxy classification further underscores the inherent ambiguity in assigning specific activity classes to galaxies around the dividing lines of a BPT diagram by showing that the class probability distributions do not align perfectly with these boundaries. This indefiniteness in the classification can be particularly important in some cases such as, for instance, when selecting Seyfert S0 galaxies, as our calculations reveal that there are zones with $P(\BPT_4) < 0.5$ above \citeauthor{Cid+2010}'s divider, especially around the intersection with \citeauthor{Kewley+2001}'s border line (top-right panel of Fig.~\ref{fig:bpt_p_bpt}). Additionally, certain regions in the uppermost part of the composite sector in the BPT diagram exhibit a probability of being classified as Seyfert with values $\gtrsim 0.5$, suggesting that some of the S0s with the highest \Oiii/\Hb\ flux ratios in this area may actually belong to this latter class. Similarly, the probability distribution for composite classification (bottom-left panel of Fig.~\ref{fig:bpt_p_bpt}) peaks in a relatively confined area near the origin of coordinates, within the lower half of this sector. Beyond this zone, the probability of being composite remains below $0.5$, dropping to values as low as $0.2$ towards the upper end of the composite region. Conversely, according to our findings the \citeauthor{Kauffmann+2003}'s border line would be the most accurate of the commonly used BPT dividers, as it traces the upper envelope of virtually all bins where the probability of a galaxy being star-forming is higher than $0.5$ and encompassing none with a probability lower than this threshold (see the top-left panel in Fig.~\ref{fig:bpt_p_bpt}). 

\begin{figure}
    \centering
	\includegraphics[width=\columnwidth]{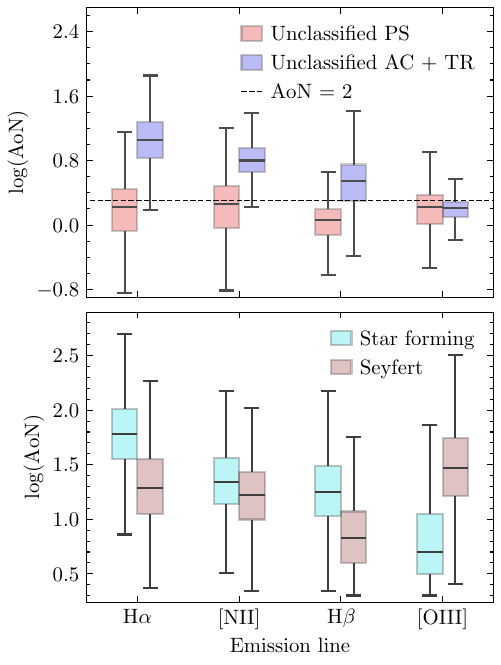}
    \caption{Box plots summarising the AoN distributions of the four diagnostic emission lines used in the BPT diagram for different subsets. \emph{Top panel:} S0-PS (red) and AC+TR galaxies (blue) that cannot be reliably classified using directly the BPT taxonomy, but for which the Portsmouth catalogue provides AoN values in the four diagnostic lines. \emph{Bottom panel:} star-forming (cyan) and Seyfert (brown) S0 galaxies. The boundaries of the boxes indicate the interquartile ranges (central 50 per cent) of the data, with the central lines signalling the median values. The horizontal dashed line included on the top panel marks the AoN threshold established by the Portsmouth emission-line flux measurements \citep{Thomas+2013} to identify reliable emission line data.}
    \label{fig:AoN}
\end{figure}

We have also demonstrated that our alternative classification scheme is applicable to galaxies with any activity level, as it relies on projections of the entire visible spectrum onto PCs, rather than focusing solely on a few significant emission lines. This makes this approach particularly valuable for studying datasets of early-type galaxies, where systems with a low activity level are prevalent. Notably, its application to a sample of nearby S0 galaxies has revealed that some systems are  excluded from BPT diagrams despite exhibiting significant activity levels (see Section~\ref{SS:unclass} and Fig.~\ref{fig:p_bpt_unclass}). The exclusion of these objects becomes evident upon examining the box plots in Fig.~\ref{fig:AoN}, which illustrate the distributions of AoN values for the four BPT diagnostic lines across specific galaxy subsets. The top panel reveals that S0s in the AC and TR spectral classes are excluded primarily because the \Oiii\ line fails to meet the minimum AoN threshold of 2  required for reliable flux measurements. Note that, as the bottom panel indicates, when star formation dominates the activity, \Oiii\ typically has the lowest AoN among the diagnostic lines. On the other hand, for AGN-driven activity, \Oiii\ exhibits the highest signal level, while \Hb\ has the lowest. It therefore follows that the uneven strength of diagnostic lines across different activity classes may introduce a bias against composite and star-forming galaxies in BPT diagrams constructed from flux-limited samples. Disregarding the important uncertainties associated with the measurement of the AoN of weak lines, the forbidden \Oiii$\lambda5007$ line also exhibits a stronger S/N than \Hb\ among the unclassified S0--PS galaxies, although it is no longer the line with the highest contrast (see the red box plots in the top panel of Fig.~\ref{fig:AoN}). These results align with the probabilities derived in the previous section from the \DPS--PC3 diagnostic. Unclassified active S0s of the AC and TR spectral classes are likely dominated by star-forming galaxies, as suggested by the relative strength of their \Oiii\ and \Hb\ lines. In contrast, for unclassified S0--PS galaxies, any low-level activity appears unrelated to SF and, despite the LINER-like features in the visual spectra of many of these systems, is also unlikely to stem from residual accretion onto a central SMBH.

Among the active objects lacking a reliable BPT classification, there is a small subset -- specifically those galaxies that appear detached from the left edge of the \DPS-PC3 diagram and exhibit the highest \DPS\ values in Fig.~\ref{fig:dps_pc3_blank} -- with AoNs for all diagnostic lines well above the required minimum. These galaxies are characterized by the presence of one or more very strong key emission lines (see examples in Fig.~\ref{fig:saturated_spectra}) to which unrealistically high flux values ($> 10^4$ units) have been assigned, despite having null warning flags. We suspect that these measurements might reflect rare undetected pipeline processing issues. Although the number of S0 galaxies potentially affected by these issues is negligible, ensuring no noticeable impact on the results presented in this paper, we have preferred to exclude them from the BPT classification.

\section{Conclusions}
\label{S:conclusion}
In this paper, we have introduced a novel galaxy classification tool that maps galaxies onto a diagram, \DPS-PC3, based on the projections of their visible spectra onto the first three principal components. Applied to a sample of $68,043$ nearby ($z\leq 0.1$) S0 galaxies from the SDSS Main Legacy Survey, this method has proven effective in capturing the full range of galaxy activity, including objects systematically excluded from widely used classifiers like the BPT diagram due to insufficiently strong emission lines. This makes the \DPS-PC3 diagnostic particularly valuable for classifying datasets of early type galaxies, where low-activity systems are prevalent.

Kernel density estimates of BPT-classified star-forming, Seyfert, composite and LINER S0 galaxies in the \DPS-PC3 diagram have been used to determine the probability that a galaxy belongs to each BPT activity class. By mapping these probabilities back to the BPT diagram, we have confirmed that our classification scheme aligns very well with the determination of the main ionization source made by the BPT classification scheme using line ratios, with class probabilities exceeding $0.5$ in much of the corresponding regions. However, the stochastic nature of our approach has also revealed that such probabilities can be relatively low near class demarcation boundaries, indicating that the classification of galaxies in these transition zones may be ambiguous. Particular caution is advised when selecting Seyfert S0s near the intersection of \citeauthor{Kewley+2001}'s (\citeyear{Kewley+2001}) and \citeauthor{Cid+2010}'s (\citeyear{Cid+2010}) dividing lines, as the probability of classification as Seyfert drops below $0.5$ in these areas. Similarly, the regions where the probability of an S0 being classified as LINER or composite exceeds $0.5$ are smaller than the areas delineated by their associated borders, especially for the latter designation. In fact, the composite probability stays below $0.5$ in the upper half of its sector, dipping to as low as $0.2$ at the upper end, where the two wings of the diagram diverge. In contrast, the probability of being star-forming consistently exceeds $0.5$ across nearly the entire region delineated by the \citeauthor{Kauffmann+2003}'s (\citeyear{Kauffmann+2003}) boundary. 

The comprehensiveness of our classification scheme has also enabled us to infer the most likely ionizing mechanism for $4640$ active S0 galaxies, where reliable BPT flux ratios could not be determined due to low S/N in at least one diagnostic line or issues with line measurements. Our methodology suggests that the majority of these galaxies ($\sim 97$ per cent in total) are likely composite or star-forming objects with relatively weak \Oiii\ emission. This finding raises concerns about a potential bias against these two activity types in emission-line-based classifications derived from flux-limited samples. Additionally, by applying our diagnostic tool to the visual spectra of S0 galaxies excluded from traditional BPT diagrams due to low activity levels, we have found that $\sim 94\%$ of these systems exhibit spectra predominantly resembling those of LINER galaxies. This significantly higher prevalence of LINER-like features compared to the active population \citep[e.g.][]{JimP+22}, coupled with the inferred behaviour of the AoN distributions for their diagnostic emission lines, suggests that the remaining activity in many of these galaxies might be driven by physical mechanisms unrelated nonetheless to black hole accretion or SF.

Another advantage of using principal components of visible spectra for galaxy classification is that it allows one to operate in a subspace where variance is maximized along orthogonal axes, something that is not inherently guaranteed when using traditional variables, such as certain line ratios or fluxes, regardless of the physical basis of their discriminating power. However, this benefit comes at the cost of these principal components having potentially complicated physical interpretations, since they may not directly correspond to easily recognizable or intuitive physical properties. Therefore, the choice of one type of classifier or another will ultimately depend on the specific objectives of the analysis.

\section*{Acknowledgements}
We acknowledge financial support from the Spanish state agency MCIN/AEI/10.13039/501100011033 and by 'ERDF A way of making Europe' funds through research grants PID2019–106027GB–C43, PID2022\-140871NB\-C21, and PID2022-140871NB-C22. MCIN/AEI/10.13039/501100011033 has also provided additional support through the Centre of Excellence Severo Ochoa's award for the Instituto de Astrof\'\i sica de Andaluc\'\i a under contract  CEX2021-001131-S and the Centre of Excellence Mar\'\i a de Maeztu's award for the Institut de Ci\`encies del Cosmos at the Universitat de Barcelona under contract CEX2019-000918-M. J.L.T.\ acknowledges support by the PRE2020-091838 grant from MCIN/AEI/10.13039/501100011033 and 'ESF Investing in your future', and by the Science and Technology Facilities Council (STFC) of the UK Research and Innovation via grant reference ST/Y002644/1.

\section*{Data availability}
The data generated in this article are available in the article and in its online supplementary material, or at the CDS website (\url{pending}). This research has made prominent use of the following databases in the public domain: the SDSS Science Archive Server (\url{https://data.sdss.org/sas/}), and the VizieR Online Data Catalog J/MNRAS/515/3956 (\url{https://cdsarc.cds.unistra.fr/viz-bin/cat/J/MNRAS/515/3956}).


\bibliographystyle{mnras}
\bibliography{biblio}



\appendix
\section{Ancillary material}
\label{A:extra_material}
This appendix provides a sample table containing key measurements related to the \DPS-PC3 and BPT diagrams, derived from the full visible SDSS spectra of the nearby S0 galaxies analysed in this study, along with additional figures that visually complement the discussions in the main text.

\begin{figure}
    \centering
	\includegraphics[width=\columnwidth]{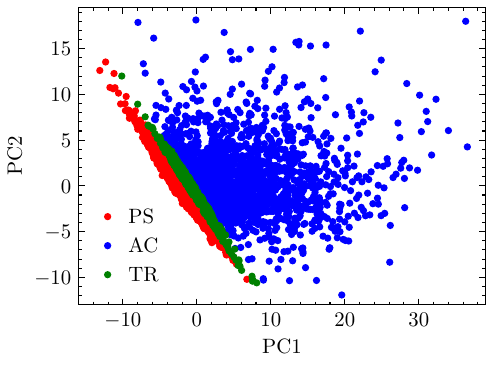}
    \caption{Spectra of present-day S0 galaxies projected onto the first two PCs. Each point represents an individual galaxy, colour-coded by its spectral class: passive sequence (red), active cloud (blue), and transition region (green). To reduce overcrowding, only a random 10 per cent of the original sample is displayed.}
    \label{fig:pc1_pc2}
\end{figure}

\begin{figure}
    \centering
	\includegraphics[width=\columnwidth]{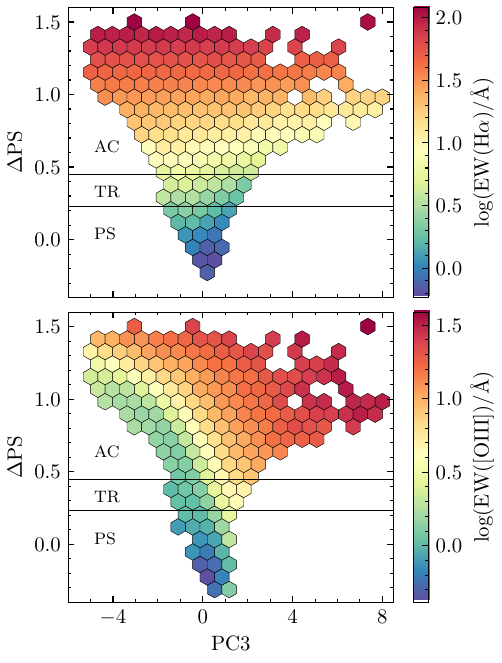}
    \caption{Same as in Fig.~\ref{fig:dps_pc3_bpx_bpy} but for the EW(\Ha) (top) and EW(\Oiii) (bottom).}
    \label{fig:dps_pc3_ssfr_oiii}
\end{figure}

\begin{figure}
    \centering
	\includegraphics[width=\columnwidth]{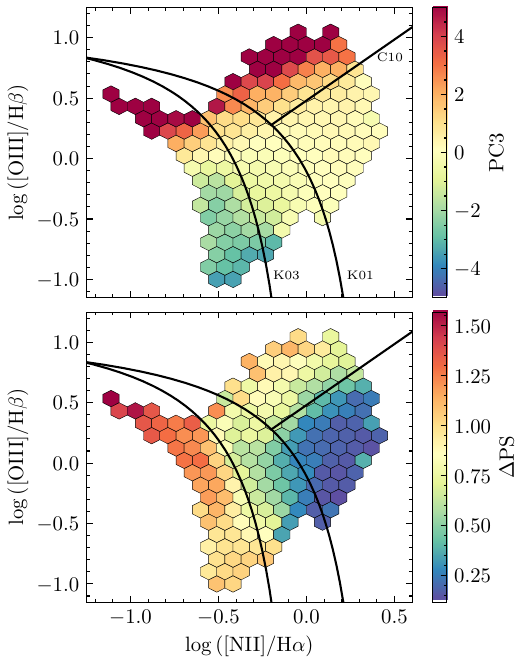}
    \caption{Distribution of PC3 (top) and \DPS\ (bottom) in the BPT diagram. The colours indicate the median value of these parameters within each hexagonal cell. Black curves represent standard class boundaries: K01 is the demarcation proposed by \citet{Kewley+2001}, K03 by \citet{Kauffmann+2003}, and C10 by \citet{Cid+2010}. Only galaxies with spectra having AoN$\,> 2$ in the four diagnostic emission lines are included in this figure.}
    \label{fig:bpt_cmap}
\end{figure}

\begin{figure}
    \centering	\includegraphics[width=\columnwidth]{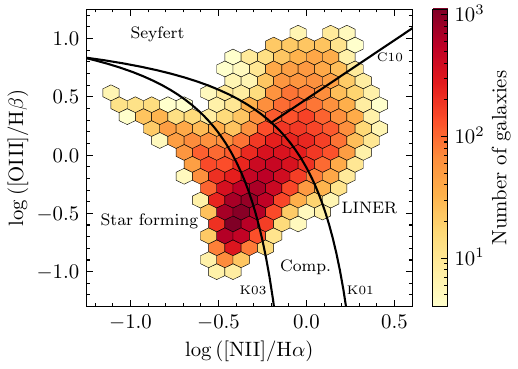}
    \caption{BPT diagram of nearby S0 galaxies with an AoN$\,> 2$ for all relevant emission lines. Galaxies are binned into hexagonal cells, with each cell colour-coded according to the number of galaxies it contains. The diagram reveals that most present-day active S0s fall into the star-forming and composite classes. The class dividers are the same as in Fig.~\ref{fig:bpt_cmap}.}
    \label{fig:bpt_hist}
\end{figure}

\begin{figure}
    \centering
	\includegraphics[width=0.8\columnwidth]{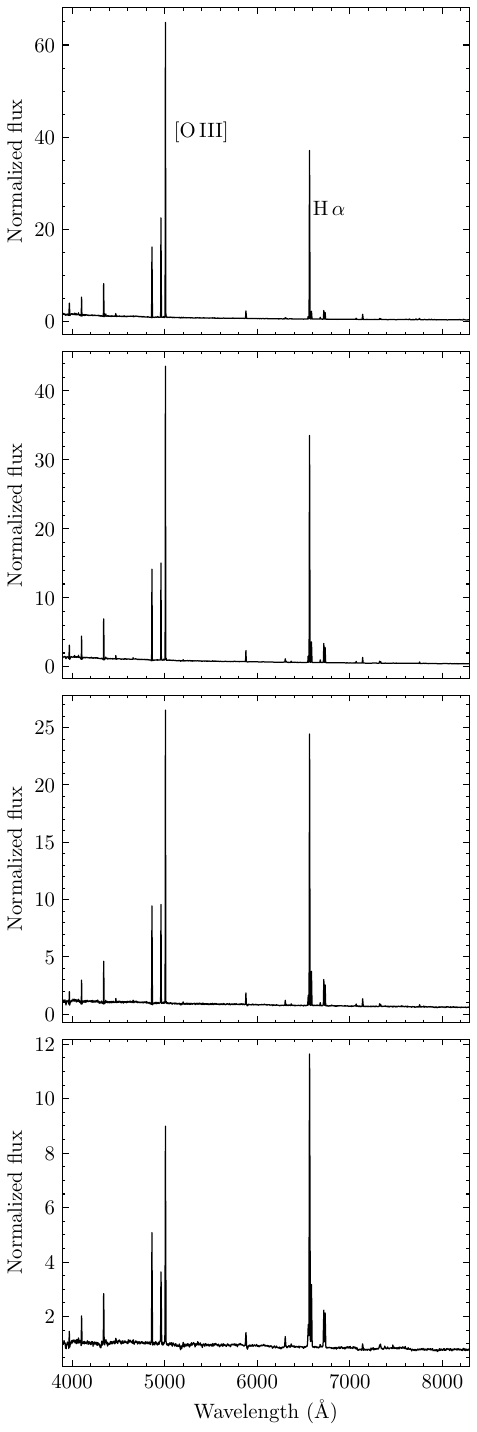}
    \caption{Four examples of S0 galaxies with strong, high S/N emission lines for which the Portsmouth catalogue assigns unrealistically high flux values ($> 10^4$ units) despite the absence of any warning flag. These galaxies have been excluded from the BPT classification. The galaxies' SDSS \texttt{ObjID}s, from top to bottom, are: $1237653650763808807$, $1237664879416049669$, $1237667783899152490$, and $1237667736664932397$.}
    \label{fig:saturated_spectra}
\end{figure}

\begin{table*}
 \caption{$\Delta$PS-PC3 coordinates, spectral class and probability of BPT type for SDSS S0 galaxies with $z\leq 0.1$.}
    \label{tab:catalogue}
\resizebox{\textwidth}{!}{    
\begin{tabular}{ccccccccc}
\hline
ObjID & PC3 & $\Delta$PS & Spectral class & $P$(star-forming)  & $P$(composite)  & $P$(LINER) & $P$(Seyfert) & BPT class$^a$ \\ \hline
1237661852543418420 & 0.4027  & -0.1465 & PS & 0.0003 & 0.1024 & 0.8882 & 0.0091 & Unclassified \\
1237662195614351581 & 5.7649  & 0.9158  & AC & 0.0001 & 0.0245 & 0.0    & 0.9753 & Seyfert      \\
1237652616740536436 & -1.1021 & 0.7651  & AC & 0.5164 & 0.4566 & 0.0221 & 0.005  & Starforming  \\
1237650370488697036 & 0.2915  & -0.1228 & PS & 0.0006 & 0.0846 & 0.9058 & 0.009  & Unclassified \\
1237668289624998066 & -0.2787 & 0.6196  & AC & 0.3053 & 0.5763 & 0.0946 & 0.0237 & Composite    \\ \hline
\end{tabular}}
    \begin{tablenotes}\footnotesize
    \item[] \emph{Note.} ($^a$) Activity class inferred from the traditional BPT-\ion{N}{ii} diagram. Probabilities were derived from eq.~\eqref{eq:p_bpt} assuming $P({\rm BPT_0}) = 0$. This is a sample table consisting of the first 5 rows of data. A machine-readable version of the full table is available online at the CDS website.
    \end{tablenotes}
\end{table*}

\bsp	
\label{lastpage}
\end{document}